\documentstyle[aps,prl,multicol,psfig,floats]{revtex}

\begin{document}

\twocolumn[
\hsize\textwidth\columnwidth\hsize\csname@twocolumnfalse\endcsname

\title{Thermodynamic stability of folded proteins against mutations}

\author{H. J. Bussemaker, D. Thirumalai, and J. K. Bhattacharjee}
\address{Institute for Physical Science and Technology,
         University of Maryland, College Park MD 20742}

\date{to be published in Phys.\ Rev.\ Lett.}

\maketitle

\begin{abstract}
By balancing the average energy gap with its typical change 
due to mutations for protein-like heteropolymers with $M$ residues,
we show that native states are unstable to mutations on a scale 
$M^\star\sim(\lambda/\sigma_\mu)^{1/\zeta_{\rm s}}$,
where $\lambda$ is the dispersion in the interaction free energies and 
$\sigma_{\!\mu}$ their typical change.
Theoretical bounds and numerical estimates (based on complete enumeration 
on four lattices) of the instability exponent $\zeta_s$ are given.
Our analysis suggests that a limiting size of single-domain proteins
should exist, and leads to the prediction that small proteins
are insensitive to random mutations.
\end{abstract}
]

\newpage

In order for a protein to perform a biological activity it has to fold
to a structure with a well-defined topology.
It is widely believed that this native state has the lowest free 
energy \cite{Anfinsen}.
It is less well appreciated that the folded state of a single-domain 
protein is only marginally stable compared to the ensemble 
of unfolded structures~\cite{newref2}.
Thus the polypeptide chain can undergo large amplitude fluctuations
after reaching the native conformation.
Another characteristic of single-domain proteins is that they are small: 
the number of amino acid residues in single-domain proteins seldom
exceeds 200.
Larger proteins usually self-organize into multi-domain 
structures~\cite{newref2}.

The marginal stability of single-domain proteins suggests that 
polypeptide chains may be sensitive to 
mutations that alter the primary sequence of amino acids.
On the other hand, many different sequences often lead to similar 
native states, i.e., sequence homology gives rise to a class 
of folded states which are topologically similar~\cite{newref2}.
Thus, from the perspective of homologous proteins we expect that point 
mutations may have little effect on the stability of the folded states 
of proteins.
These observations raise the question: to what extent are native states of
proteins stable when subjected to random mutations?

Experimentally, the stability of folded states against random uncorrelated 
mutations can only be examined by synthesizing a library of mutated sequences 
starting from a known structure~\cite{Davidson95}.
In this article we investigate the stability of folded states using a 
combination of theoretical arguments and exact enumeration for various lattice
models of proteins.

Since our arguments are general, we can only quantify the effect of mutations 
in statistical terms.  In particular, we have analyzed how the probability 
that a mutation causes a change in the native state of a protein depends on 
the number of residues.
We assume that the conformations in the native and first excited state are
not structurally related, so that it is reasonable to focus on 
these two states to estimate the effect of random mutations.
For the simpified models discussed below we have verified this assumption
by explicit computation of the overlap between the two states for a number
of sequences.
For real proteins the gap would correspond to the free energy difference
between the native basin of attraction (NBA) and the closest competing basin 
of attraction (CBA);  the structures in the NBA and CBA are expected to be
dissimilar as well.

The main result of our analysis is that the energy gap $\Delta_{\rm av}$, 
averaged over a large number of sequences, and the typical change 
$\sigma_{\!\Delta}$ in the energy gap due to mutations scale with the number 
of residues $M$ as
\begin{equation}\label{scaling}
	\Delta_{\rm av} \sim \lambda M^{-\theta},
	\qquad 
	\sigma_{\!\Delta} \sim \sigma_{\!\mu} M^\alpha, 
\end{equation}
with $\alpha,\theta>0$, where $\lambda$ is the dispersion of the interaction 
free energies along the chain, and $\sigma_{\!\mu}$ is the typical change in 
interaction free energy for each residue due to mutations.
Since $\Delta_{\rm av}$ decreases with $M$ while $\sigma_{\!\Delta}$ increases,
it follows that there is a crossover length $M^\star$, scaling with the
instability exponent $\zeta_{\rm s}=\alpha+\theta$ as
\begin{equation}\label{M-star}
	M^\star \sim (\lambda/\sigma_{\!\mu})^{1/\zeta_{\rm s}}.
\end{equation}
The native state is robust with respect to mutations if 
$M\ll M^\star$, but there is a considerable chance that the native state 
is destabilized by mutations if $M \agt M^\star$.

An estimate for $\alpha$ defined in
Eq.~(\ref{scaling}) goes as follows.
Let $n_{\!\Delta}$ denote the number of contacts that are different between 
the ground state and the first excited state.
The independent mutations in the interaction free energies, of typical size
$\sigma_\mu$ for each residue, combine to give a change in the gap energy of
magnitude $\sigma_{\!\Delta}\sim\sigma_{\!\mu}\sqrt{n_{\!\Delta}}$.
Assuming that the $n_\Delta$ different contacts are located on a fractal
surface, we obtain the inequality $\alpha \geq (d-1)/2d$.
The maximum possible number of contacts, which can only occur for $d=\infty$,
is $n_{\!\Delta}^{\rm max}=M^2$.  Thus $(d-1)/2d\leq\alpha<1$.

To motivate why $\Delta_{\rm av}$ decreases algebraically with $M$
consider an ensemble of $n$ random energies, drawn from a Gaussian 
distribution $f(E)\propto\exp(-E^2/2)$.
The distribution function for the energy gap $\Delta$ between the two smallest 
values among the $n$ samples is given by
$f(\Delta) \propto n \int_{-\infty}^\infty dE\,f(E)\,[g(E+\Delta)]^{n-1}$,
where $g(E)=\int_E^\infty dE' f(E')$ equals the probability that the
energy is larger than $E$.
It turns out (see also below) that $f(\Delta)$ is very well described by an 
exponential distribution function,
$f(\Delta) = (1/\Delta_{\rm av}) \exp(-\Delta/\Delta_{\rm av})$.
Consequently, we have 
$1/\Delta_{\rm av}=-(d/d\Delta)\log f(\Delta)|_{\Delta=0}%
=-n\int_{-\infty}^\infty dE\,E\,f(E)\,[g(E)]^{n-1}$.
By numerically evaluating this integral for various values of $n$ we find 
that for large $n$ the average gap decreases as 
$\Delta_{\rm av} \sim (\log n)^{-\theta}$ with $\theta\simeq0.2$ in the 
range $10 \leq n \leq 10^6$.
Since different conformations may have considerable overlap, the effective 
number $n$ of independent samples is smaller than the total number of 
conformations, which grows exponentially with $M$.
But it is reasonable to assume that $n$ still depends exponentially on 
the number of residues, so that $n\sim c^M$ gives rise to 
$\Delta_{\rm av} \sim M^{-\theta}$.
This power law behavior is independent of the constant $c$.

To numerically verify the scaling behavior of Eq.~(\ref{scaling}), which leads 
to the length scale $M^\star$ (cf.\ Eq.~(\ref{M-star})), we have used 
simplified lattice models for proteins \cite{Dill95} that allow for complete 
enumeration of all conformations for chains of moderate length.
In these models, a protein is re\-presented by a self-avoiding walk on a 
lattice.
Natural proteins are made from twenty amino acids, roughly half of which
are hydrophobic.
Contacts between hydrophobic residues which form the core of the protein 
are favored.
To mimic the diversity of natural proteins we use a model in which each 
residue is assigned a random ``charge'', $\lambda_j$, drawn from a 
Gaussian distribution with average $\langle \lambda_j \rangle = \lambda_0$ 
and variance $\sqrt{\langle (\lambda_j-\lambda_0)^2 \rangle} = \lambda$
\cite{Garel94}.
Positive $\lambda_j$'s may be identified as hydrophilic, whereas negative
values of $\lambda_j$ correspond to hydrophobic residues.
As a typical value for the ratio of the average hydrophobicity and its 
dispersion we use $\lambda_0/\lambda = -0.1$.
This choice is consistent with the fact that in nature approximately 55\% of 
the residues are hydrophobic \cite{percentage}.
We have used this value in most of our calculations, but as will be shown 
below, the results are rather insensitive to $\lambda_0$ in the range
$-1 < \lambda_0/\lambda < 0$.

We use four different lattices: the square and the triangular lattice in 
two dimensions, and the diamond and the cubic lattice in three dimensions.
Consider a conformation $\nu$ in which the $M$ residues in the chain
reside on lattice sites ${\bf r}^\nu_j$, with $j$ labeled $1$ through $M$.
By definition, a contact occurs between two residues $j$ and $k$ if they are 
nearest neighbors on the lattice.
In terms of a (symmetric) contact matrix $C$ a contact is represented by
$C^\nu_{jk}=1$.
Subsequent residues $j$ and $j+1$ are always nearest neighbors on the 
lattice, and therefore $C^\nu_{j,j+1}\equiv0$.
We also define $C^\nu_{jj}\equiv0$.

When two residues $j$ and $k$ are in contact, they give a contribution to the
total energy $E^\nu$ that is by definition equal to the average, 
$E_{jk}=(\lambda_j+\lambda_k)/2$, of their ``charges''.
Thus, the total energy of a conformation is given by
\begin{equation}\label{energy}
	E^\nu 
	= \sum_{j<k} C^\nu_{jk} \left(\frac{\lambda_j+\lambda_k}{2}\right)
	= \sum_j n^\nu_j \left(\frac{\lambda_j}{2}\right),
\end{equation}
where $n^\nu_j=\sum_{k\neq j}C^\nu_{jk}$ equals the number of contacts in 
which residue $j$ is involved.
For end point residues, $n_j$ takes values from $0$ to $z-1$, where $z$ is
the coordination number of the lattice; for internal residues the maximum 
number of contacts equals $z-2$.

It is important, both from a theoretical and a computational point of
view, that the energy of a conformation $\nu$ depends on the coordinates 
$\{{\bf r}^\nu_j\}$ only through the contact matrix $\{C^\nu_{jk}\}$,
and in fact an even more reduced description in terms of $\{n^\nu_j\}$ 
is possible.
Therefore, as long as we are interested in thermodynamic rather than kinetic
properties, it is sufficient to consider all possible contact sets 
$\{n^\nu_j\}$.
For each value of $M$ we enumerated all possible conformations 
$\{{\bf r}^\nu_j\}$, determining $\{n^\nu_j\}$ for each conformation, 
and only adding it to the list if it is different from all previously
generated contact sets.

The list thus generated was used to find the energies of the native and the
first excited state for a large number of random sequences.
Many different spatial conformations $\{{\bf r}^\nu_j\}$ can give rise 
to the same set of pair contacts $\{n^\nu_j\}$, which makes our approach 
very efficient.
To quantify this, we first note that the total number of different 
conformations, $C_M$, grows exponentially with $M$, 
as $C_M\sim z_{\rm eff}^M$, where $z_{\rm eff}$ is an effective coordination 
number \cite{Guttmann}.
The number of contact sets $C'_M$ also grows exponentially with $M$, but with
an effective coordination number $z'_{\rm eff}<z_{\rm eff}$, 
so that $C'_M/C_M\sim(z'_{\rm eff}/z_{\rm eff})^M$.
On the cubic lattice $z_{\rm eff}\simeq4.684$ \cite{Guttmann}
while $z'_{\rm eff}\simeq3.4$, and for $M=15$
the contact set enumeration is already more efficient by a factor of 400.
The description of the energies in terms of the variables $\{n^\nu_j\}$
therefore allows us to calculate the low energy states for a much larger 
number of sequences than would otherwise have been possible.

\begin{figure}
\psfig{file=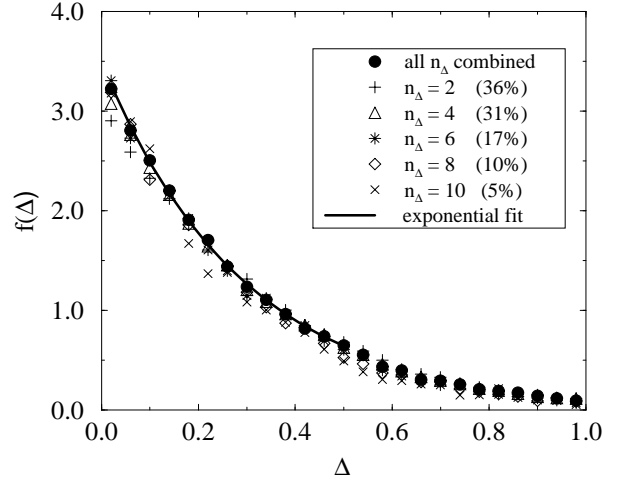,width=\columnwidth}
\caption{
Distribution of energy gap $\Delta$ between native and first excited state 
for the diamond lattice model with $\lambda_0=-0.1$ and $M=18$,
obtained using an ensemble of $10^5$ sequences.
The solid line denotes an exponential fit to the data shown as filled circles.
The other symbols correspond to a restriction to specific values of 
$n_{\!\Delta}$.
The percentages indicate the fraction of sequences for which each
$n_{\!\Delta}$ occurs.
}
\label{fig1}
\end{figure}

Consider the ground state ($\nu=0$) and first excited state ($\nu=1$)
for a given sequence.
The energy gap, $\Delta=E^1-E^0$, is defined as the energy difference
between the native state and the first excited state.
A histogram of $\Delta$, obtained by determining the native state and 
first excited state for $10^5$ randomly drawn sequences $\{\lambda_j\}$ fits 
very well to the exponential distribution (see Fig.~\ref{fig1}).

The mutations are modeled by adding a small perturbation to the ``charges'',
$\lambda_j \to \lambda_j + \mu_j$,
where $\mu_j$ is a Gaussian random variable with zero mean and variance
$\sigma_{\!\mu}$.
Other ways of simulating the effect of mutations in lattice models have 
also recently been considered~\cite{Banavar}.
The effect of a mutation depends on the degree of overlap between 
$\{n^0_j\}$ and $\{n^1_j\}$.
Mutations affecting contacts that occur in both the native state and 
the first excited state have no influence on the energy gap $\Delta$, since 
$E^0$ and $E^1$ are changed by the same amount.
When a contact is not shared between the two states however, mutations will
affect the energies of both states differently.
To quantify the effect of the mutations in terms of $\{n^0_j\}$ and $\{n^1_j\}$
we consider the effective number of different contacts,
\begin{equation}
	n_\Delta = \sum_{j=1}^M\ (n^1_j-n^0_j)^2.
\end{equation}
As a result of the mutations the energy gap is changed 
by an amount equal to the sum of $n_{\!\Delta}$ independent Gaussian
variables $\mu_j/2$, $\Delta \to \Delta + \mu_{\!\Delta}$,
where $\mu_{\!\Delta}$ is a Gaussian variable with zero mean
(this is a direct consequence of the fact that $\langle\mu_j\rangle=0$)
and variance $\sigma_{\!\Delta}=\sigma_{\!\mu}\langle\sqrt{n_\Delta}\rangle/2$.
The values of $n^2_\Delta$ for an ensemble of sequences $\{\lambda_j\}$ are 
exponentially distributed for large enough $M$.

Since $\Delta$ and $\sigma_{\!\Delta}$ are both derived from the same native 
state and first excited state, it is important to analyze to what extent 
they are correlated.
To this end we have plotted in Fig.~\ref{fig1} the distribution of the energy
gap $\Delta$ for subsets of sequences for which $n_\Delta$ has a particular 
value.
It can be seen that the distribution is essentially independent of $n_\Delta$.
Therefore we are justified in considering $\Delta$ and $n_{\!\Delta}$,
or equivalently $\Delta$ and $\sigma_{\!\Delta}$, to be independent.

\begin{figure}
\psfig{file=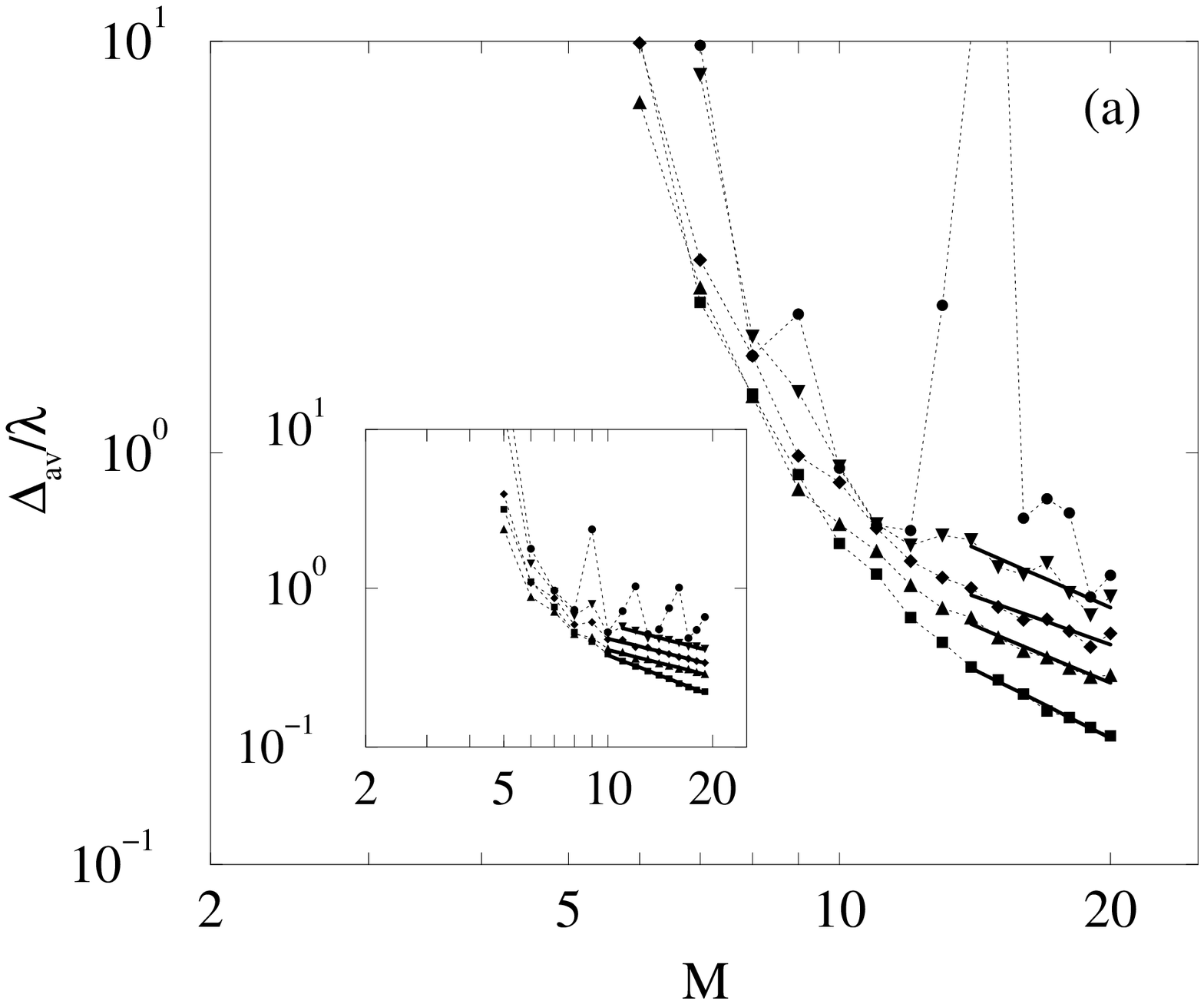,width=\columnwidth}
\psfig{file=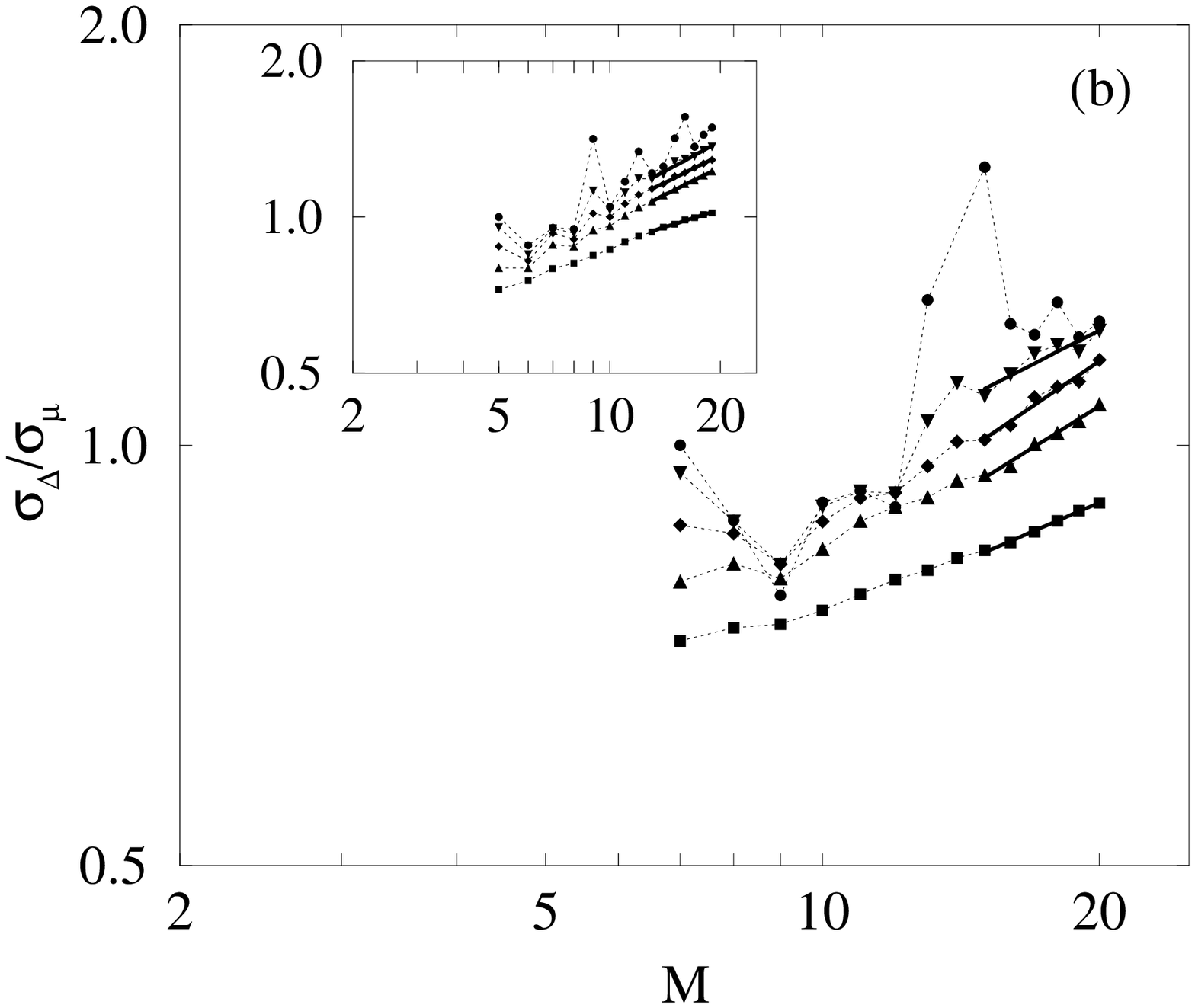,width=\columnwidth}
\caption{
Dependence of 
(a) the average energy gap $\Delta_{\rm av}$, and 
(b) the typical change in the gap due to mutations $\sigma_{\!\Delta}$, 
on the number of residues $M$ for the diamond lattice model. 
The insets show the same data for the square lattice model.
Different symbols denote various values of $\lambda_0/\lambda$, from top to
bottom: $-\infty$, -1.0, -0.5, -0.1, and 0.5.
The data were obtained by averaging over $40,000$ random sequences.
}
\label{fig2}
\end{figure}

Figure~\ref{fig2} shows how $\Delta_{\rm av}$ and $\sigma_{\!\Delta}$ 
depend on the number of residues for the square and the diamond lattice models:
$\sigma_{\!\Delta}/\sigma_{\!\mu}$ is an increasing function of $M$, 
while $\Delta_{\rm av}/\lambda$ decreases with $M$.
This behavior holds qualitatively for all four lattices studied.
Power law fits to the data for large $M$, shown as solid lines in 
Fig.~\ref{fig2}, provide estimates for the exponents $\alpha$ and $\theta$ 
defined in Eq.~(\ref{scaling}) for various values of $\lambda_0/\lambda$.
In the range $-0.5<\lambda_0/\lambda<0$ there is no significant dependence
on $\lambda_0/\lambda$.
For $\lambda_0/\lambda=0.5$, when the chain is largely hydrophilic,
$\alpha$ and $\theta$ have different values, although their sum 
(and consequently the instability exponent $\zeta_{\rm s}=\alpha+\theta$)
hardly changes.
However, in the limit $\lambda_0/\lambda\to-\infty$, where only maximally 
compact states contribute, the scaling behavior breaks down completely, and the
dependence on $M$ is quite erratic for the whole range of $M$ values considered.

On all four lattices we find $\alpha=0.3\pm0.1$, which is consistent with 
the theoretical bound $(d-1)/2d\leq\alpha<1$ in both two and three dimensions.
We find $\theta=0.6\pm0.2$ in two dimensions (square and triangular lattice)
and $\theta=1.0\pm0.2$ in three dimensions (diamond and cubic lattice).
Both these values are significantly larger than the theoretical prediction 
$\theta\simeq0.2$. 
It is quite possible that in the range of $M$ values that allow for complete
enumeration, $\theta$ is enhanced by finite-chain-length effects,
and that for larger $M$ the decay of $\Delta_{\rm av}/\lambda$ is slower 
and $\theta$ reaches its predicted value $\theta\simeq0.2$.
But for $M^\star$ as defined in Eq.~(\ref{M-star}) to exist we only
require that $\theta>0$.
The values of the instability exponent $\zeta_{\rm s}=\alpha+\theta$ for
the various lattices are
$0.9 \pm 0.1$ (square), $1.0 \pm 0.2$ (triangular),
$1.2 \pm 0.2$ (diamond), and $1.3 \pm 0.3$ (cubic).
A more accurate determination of the exponents would require much larger
values of $M$.

In Ref.~\cite{NEC} it was shown that an energy function like
the one defined in Eq.~(\ref{energy}) ---
i.e.\ a sum of one-residue properties rather than a sum of pair interactions
--- provides a good parametrization of the interaction matrix that was
constructed by Miyazawa and Jernigan \cite{MJ} by statistically analyzing
a large number of known protein structures obtained from the Protein Data
Bank.
To verify this claim explicitly we have performed calculations using the 
interaction matrix of Ref.~\cite{MJ} and find values for the exponents that 
are in the same approximate range as the values obtained using the ``charge'' 
model \cite{Bussemaker}.

In obtaining the above results for random heteropolymers we have averaged over 
all sequences without regard to the degeneracy of the ground state.
Since the native states of proteins are thought to be unique \cite{Anfinsen},
we have repeated our calculations by including only those sequences that
have a non-degenerate native state.
Surprisingly, the values of $\alpha$ and $\theta$, and hence of the instability
exponent $\zeta_{\rm s}$, are virtually unchanged.

It might be argued that for our results to be applicable to proteins,
which in the course of evolution may have been ``optimized'' (e.g.\ with
respect to folding kinetics \cite{optimal}),
one must only include ``optimized'' sequences in the averaging process.
Although this may change the value of the instability exponent $\zeta_{\rm s}$,
we expect that $\alpha$ and $\theta$ will both remain positive.
This expectation is reasonable given the fact that our results are virtually
unchanged when we consider only non-degenerate ground states, as mentioned
above.
Thus we expect our conclusion that the native state is unstable against weak 
mutations for sufficiently long chains also to hold for (single-domain) 
proteins.

Since in biological systems mutations occur at a given rate and the dispersion
in the interaction free energies of the residues has a limited range of values
\cite{newref2},
the fact that single-domain proteins with more than about 200 residues rarely 
occur may be explained by the sensitivity of the native state to small random
perturbations when the number of residues is large.
To make this more precise, consider 
$p_{\rm d} = \int_0^\infty d\sigma \int_0^\sigma d\Delta
f(\Delta|\Delta_{\rm av})g(\sigma|\sigma_{\!\Delta}),$
i.e., the probability that the folded state is destabilized by the mutations, 
which only depends on the ratio $x\equiv\sigma_{\!\Delta}/\Delta_{\rm av}$.
Here $f(\Delta|\Delta_{\rm av})$ is an exponential distribution with average
$\Delta_{\rm av}$ and $g(\sigma|\sigma_{\!\Delta})$ is a Gaussian distribution 
with zero mean and width $\sigma_{\!\Delta}$.
For small $x$ we have $p_{\rm d}\sim x$ so that using Eq.~(\ref{M-star})
we obtain $p_{\rm d} = c(M/M^\star)^{\zeta_{\rm s}}$ for $M\ll M^\star$,
with $c$ a constant.
Now let $\varepsilon$ be a small threshold probability separating
stable ($M<M_\varepsilon$) from unstable ($M>M_\varepsilon$) proteins,
where $M_\varepsilon$ is defined as the value of $M$ for which 
$p_{\rm d}=\varepsilon$.
Eliminating $M^\star$ in favor of $M_\varepsilon$ we obtain
\begin{equation}\label{p_d}
	\frac{p_{\rm d}}{\varepsilon}
	=\left(\frac{M}{M_\varepsilon}\right)^{\zeta_{\rm s}}.
\end{equation}
Given a pair ($\varepsilon,M_\varepsilon$) we can therefore predict the 
stability for other chain lengths $M\alt M_\varepsilon$ using Eq.~(\ref{p_d}), 
provided that we know the value of the instability exponent $\zeta_{\rm s}$.
Note that the ratio of the $p_{\rm d}$ values for two different values of $M$ 
is independent of $\varepsilon$, and is thus a suitable quantity to be 
considered in experiments designed to verify Eq.~(\ref{p_d}).
Since $\zeta_{\rm s}\simeq 1$, the destabilization probability $p_{\rm d}$
decreases more or less linearly with $M$.
It follows that on average small proteins are relatively insensitive to point 
mutations.

In order to utilize our theory for analyzing experiments, it should be kept 
in mind that the models provide a coarse-grained 
description of a protein, where each bead corresponds to $b$ amino acids, 
with $3\alt b\alt 4$ \cite{coarse}.
In terms of the number of amino acids $M_{\rm aa}$ and the free energy 
dispersion per amino acid $\lambda_{\rm aa}$, the corresponding model 
parameters are $M=M_{\rm aa}/b$ and $\lambda=\lambda_{\rm aa}\sqrt{b}$.
The coarse-graining does not affect Eq.~(\ref{p_d}).

It is important to estimate the effect of thermal fluctuations on the
predictions of our theory.
According to Eq.~(\ref{scaling}),
for large enough $M$ we have $\Delta_{\rm av}=c_\Delta\lambda M^{-\theta}$, 
with 
$c_\Delta\simeq 3$ and $\theta\simeq1$ obtained from the power law fits of 
Fig.~\ref{fig2}.
Taking into account the coarse-graining factor $b$ and estimating 
$\lambda_{\rm aa} \simeq (15$-$25)\,kT$, 
we find for medium-sized proteins with $M_{\rm aa}\simeq100$ that 
$\Delta_{\rm av}\simeq (3$-$5)\,kT$.
We conclude that thermal fluctuations do not significantly affect 
our conclusions as summarized by Eq.~(\ref{M-star}).
Note that the scaling of $\Delta_{\rm av}$ with $M$ suggests that smaller 
proteins in general have larger melting and denaturation temperatures 
\cite{Alexander}.

There are some striking similarities between our findings and certain results 
probing the chaotic nature of the spin-glass phase.
Using an Imry-Ma style argument \cite{Imry+Ma}, Bray and Moore \cite{Bray+Moore}
showed that in the ground state of a disordered Ising model the average 
excitation energy for a cluster of linear size $L$ scales as $L^y$,
while the typical energy associated with random bond perturbations scales as
$L^{d_{\rm S}/2}$, with $d_{\rm S}/2>y$.
As a consequence, at sufficiently large length scales the ground state is 
unstable against arbitrarily weak perturbations to the bonds.
In this sense, the ground state of spin-glasses is ``chaotic''
\cite{Bray+Moore,McKay82,Derrida83}, 
and by analogy the same could be said about the native state of a protein.

One of us (HJB) acknowledges stimulating discussions with M.~Betancourt, 
D.~Klimov, and A.~Latz.
This research was performed under NSF Grant No.\ CHE96-29845.

\end{document}